\documentclass{rspublic}
\usepackage{graphicx} 
\usepackage{subfigure} 
\usepackage{t1enc,natbib,color}
\usepackage{color}

\newcommand{\zetat}{\tilde{\zeta}}

\begin{document}

\title{Determination of the zeta potential for highly charged colloidal suspensions}

\author{Giovanni Giupponi and Ignacio Pagonabarraga}
\affiliation{Departament de Fisica Fonamental, Universitat de Barcelona, Carrer Marti i Franques, 08028 Barcelona (Spain)}

\maketitle

\begin{abstract}{Zeta-potential, electrophoresis, lattice-Boltzmann, mesoscopic models, electrostatic radius.}
We compute the electrostatic potential at the surface, or zeta potential
$\zeta$, of a charged particle embedded in a colloidal suspension using a
hybrid mesoscopic model. We show that for weakly perturbing electric
fields, the value of $\zeta$ obtained at steady state during
electrophoresis is statistically indistinguishable from $\zeta$ in
thermodynamic equilibrium. We quantify the effect  of  counterions
concentration on $\zeta$. We also evaluate the relevance of the lattice
resolution for the calculation of $\zeta$ and discuss how to identify the
effective electrostatic radius.
\end{abstract}

\section{Introduction}
\label{Introduction} 

A quantitative understanding of the electrostatic interactions
between charged macroions in solution is fundamental to comprehend a
plethora of physical phenomena spanning from biology to material science, for
example the macroscopic and rheological properties of colloidal
suspensions[1,2]. The details of such interactions are
difficult to capture, as the effective interactions are determined by the
interplay between the different components dissolved in solution (macroions,
counterions, salt ions) and the solvent dielectric response[3]. In addition, when
the system is driven out of equilibrium, for example by an external electric
field causing electrophoretic flow, hydrodynamic interactions between
the solvent and solute species must also be included and can
alter the equilibrium electrostatic interactions.

The simplest yet very useful model to describe electrolyte solutions is the
Poisson-Boltzmann (PB) approach[1,4], which builds on a
continuous description of the electrolyte, characterized in terms of the anion
and cation local densities. Within this framework, it is possible to add
charged macroscopic objects, with fixed surface charges, by accordingly
changing the electrostatic boundary conditions of the PB equations. Notwithstanding
PB is a mean-field theory that defines ions as point-like and therefore
neglects excluded volume effects and spatial correlation due to their
finite size, it has been successfully adopted to describe various physical
systems, for example to derive the electrostatic part of DLVO interparticle
potential[1] which explains the interactions of weakly charged
particles in solution at low volume fraction.
  
The zeta potential, $\zeta$, defined as the electrostatic potential at the colloid
surface, or slippling plane, plays a central role for charged colloidal dispersions, 
as it indirectly provides an estimate of the magnitude of the electric field
between particles that results from the combined effect of ionized charged
groups sitting on the particle, counterions released into solution and salt
ions.  Theoretically, Ohshima {\it et al.}[5] obtained an exact
analytic expression relating the surface charge density to the zeta potential
for an infinitely dilute spherical colloidal suspension by carefully
approximating the non-linear PB equation.  This expression is of limited
  use, since the infinite dilution  regime is  difficult to achieve.
  Moreover, experimentally $\zeta$ is normally derived  out of
  equilibrium, from electrophoretic mobility measurements.  Since, {\it a
    priori}, the values of $\zeta$ in equilibrium and at steady state will
    differ,  an accurate description of hydrodynamics must be included in
    order to derive the relationship between $\zeta$  and the colloid
    mobility. 

Analytical results for the electrophoretic flow at finite dilution cannot in general be obtained.
Computationally, different models have been used to describe
electrophoresis[6-9,11]. Lobaskin {\it et al.}[8] and D\"unweg {\it et al}[9],
using a lattice Boltzmann[10] (LB) solver coupled to molecular dynamics of
ions, calculate the particle mobility for low to zero salt concentration
explicitly accounting for finite ions size, but without providing results for
$\zeta$. Kim {\it et al.}[11] show agreement with Ohshima's results, but only
for weakly charged particles at low salt concentration. Cell models[12] are
also employed to describe electrokinetics and to derive mobility versus
$\zeta$ curves; however, they rely on somewhat arbitrary boundary conditions
both for electrostatics and hydrodynamics.
      
In this paper, we compute $\zeta$ using a mesoscopic model that
couples a Navier-Stokes solver to the convection and diffusion of ions on a
lattice.  We will first investigate the difference between $\zeta$ values at
thermodynamic equilibrium and at electrophoretic steady state. We then
accurately analyze the dependence of $\zeta$ on colloid volume fraction,
$\Phi$, and salt and counterions concentration, and compare them
with predictions from PB at infinite dilution derived by Ohshima
{\it et al.}[5].  Finally, we will address how to identify the electrostatic
radius of a colloid on a lattice, as it  has been done previously for the
hydrodynamics radius[13]. In the following section, we briefly introduce our
model, the reference equations and results that can be obtained using PB
equation for infinite dilution and at finite volume fractions. We present in
section \ref{Results} results for $\zeta$ as a function of the  volume
fraction, salt concentration and lattice refinement. Finally, we draw our main
conclusions in the last section.

\section{Model}
\label{Model} 

\subsection{Electrophoretic flow}
In order to derive a correct relation between the mobility and the zeta potential at
steady state during electrophoresis, our model includes hydrodynamic
interactions and electric forces due to local charge densities,
\begin{equation}
\label{eq1}
\frac{\partial \rho_k}{\partial t} = - \nabla \cdot \rho_k\vec{v} + \nabla \cdot D_k[\nabla \rho_k + e\beta z_k \rho_k \nabla \varphi],
\end{equation}
\begin{equation}
\label{eq2} 
\frac{\partial(\rho \vec{v})}{\partial t} = \eta \nabla^2 \vec{v} - \nabla p_{id}+ \beta\sum_k e z_k\rho_k \nabla \varphi,
\end{equation}
\begin{equation}
\label{eq3} 
\nabla^2 \varphi = - \frac{1}{\epsilon}\sum_k e z_k\rho_k, 
\end{equation}
where  $D_k, z_k, k=+,-$ are the diffusivities and valences of positive and
negative ions, $\rho, \vec{v}, p_{id}$ and $ \eta$ correspond to  the solvent
density, velocity, ideal pressure and shear viscosity, $e$ is the electron
charge, $\beta^{-1} = k_BT$ is the Boltzmann factor, $\varphi$ is the
electrostatic potential and  $\epsilon$ the solvent permittivity.
Eq.~\ref{eq1} expresses ions  mass conservation during diffusion and
advection, coupling ion dynamics to solvent motion. This is in turn
  described by the Navier-Stokes equation for a viscous fluid which  couples
    solvent dynamics to electrostatic forces due to local charge density,
            Eq.~\ref{eq2}. Finally, the Poisson equation enforces the
            electrostatic coupling between the
charged species and the embedded solid particles. 
We solve these electrokinetics equations combining an LB approach
for the  momentum dynamics with a numerical solver for  the discretized
convection-diffusion dynamics, Eq.~(\ref{eq1}), based on link fluxes[14]. Each lattice site is labeled as solid or
fluid and we consider here a single spherical object of radius $r=a$
embedded in a cubic lattice of volume $L^3$ with periodic boundary conditions.
Suspended particles are therefore resolved and  interact with the neighbouring
fluid through bounce-back[13]. The electrostatic potential 
is computed using a successive over-relaxation scheme
(SOR)[16].  This method has been successfully applied to analyze 
different dynamical processes involving suspensions of charged
objects[15,17,18,19].

\subsection{Poisson-Boltzmann electrostatics}
At equilibrium, using a mean-field approach that neglects excluded volume effects
and correlation between ions, PB equation reads[4]
\begin{equation}
\label{PBeq}
\nabla^2 \varphi(\vec{r}) = \frac{8\pi e z \rho_0}{\epsilon}\sinh(\frac{e z \varphi(\vec{r})}{k_B T}), 
\end{equation}
where a symmetric electrolyte $z_+ = z_- = z$ has been used for
simplicity and $\rho_0$, the uniform macroscopic counterion and coion number
concentration far from the colloid $\rho_+ = \rho_- = \rho_0$, is assumed to be
equal to that of an electrolyte reservoir with dissolved salt $c_{salt} = 2\rho_0$.

An analytical solution of eq.~(\ref{PBeq}) for a 
single particle is not available, as the PB equation can be solved analytically only 
for a few symmetrical configurations. Ohshima {\it et al.}[5] analyzed eq.~(\ref{PBeq})
around a spherical  colloid of charge $Q$ using a perturbative approach and derived a
quasi-exact analytical expression for $\zeta$,
\begin{equation}
\label{Ohshima}
Q = 2\frac{e a^2}{\lambda_D l_B}\sinh(\zetat/2)\left[ 1 + \frac{2}{(\kappa a) \cosh^2(\zetat/4)} + \frac{8\log (\cosh^2(\zetat/4))}{(\kappa a)^2 \sinh^2(\zetat / 2)} \right]^{1/2},
\end{equation}
where $\kappa = \sqrt{\frac{4\pi e^2 z^2 c_{salt}}{\epsilon k_B T}}=\lambda_D^{-1}$ is the inverse of the Debye length, $l_b = \frac{e^2}{4\pi
\epsilon k_B T}$ the Bjerrun length and $\zetat = \frac{e \zeta}{k_B T}$ the adimensional zeta
potential.

Useful global information can be obtained using the
Debye-H\"uckel approximation, i.e. linearizing eq.~(\ref{PBeq}). When $\frac{e z\varphi}{k_B T} \lesssim 1.0$,
\begin{equation}
\label{DHeq}
\nabla^2 \varphi(\vec{r}) = \frac{4\pi e^2 z^2 \rho_0}{\epsilon k_B T} \varphi(\vec{r}) = \kappa^2  \varphi(\vec{r}).
\end{equation}
from which one can derive the electrostatic potential around a charged spherical colloid
\begin{equation}
\label{DHsol}
\varphi(r) = \frac{\zeta a}{r} \exp^{(-(r-a)/\lambda_D)}.
\end{equation}
In the linear regime, a charged particle  in the presence of salt  develops a screened, Yukawa-like electrostatic potential, with a
characteristic length scale of $\lambda_D$ and strength given by the zeta
potential, $\zeta$.

Eq.(\ref{PBeq}) is strictly valid when the amounts of positive and negative
charges dissolved in solution are equal, i.e. when the counterions concentration is 
negligible. Experimentally, this can be obtained separating the colloidal particles from 
a bulk electrolyte solution with a semi-permeable membrane. For the
experimental set-ups with non-zero concentration of counterions, eq.~(\ref{DHeq})
becomes
\begin{equation}
\label{DHeq2}
\nabla^2 \varphi(r) = \kappa^2 \left[\frac{\rho_{c0}}{2\rho_{s0}} + (1 + \frac{\rho_{c0}}{2\rho_{s0}} )\varphi(r)  \right], 
\end{equation}
where $\rho_{s0}, \rho_{c0}$  are the reference salt and counterion concentrations in the reservoir.
Eq.~(\ref{DHeq2}) stresses the fact that, especially at
low salt concentration $c_s$ (low $\rho_{s0}$) and high volume fraction $\phi = \frac{4\pi a^3}{ L^3}$
(high $\rho_{c0}$), a deviation from the theoretical Debye-H\"uckel results
can be observed. 

Our aim is to measure $\zeta$ for different experimental regimes, analyze its behavior beyond the linear approximation (when  $\frac{e\varphi}{k_B T} \gtrsim 1$) and understand how to overcome the intrinsic inaccuracies in the location of the electrostatic radius of any lattice model. In the LB approach, we will refer to $\zeta$ as the average electrostatic potential between lattice
points pairs (boundary links) $l$ that link a colloid and a fluid node
calculated half-way between the two nodes using linear approximation,
\begin{equation}
\label{zetaMP}
\zeta_{MP} =  \left<  \sum_{l} \frac{\varphi_{solid}(l) + \varphi_{fluid}(l)}{2} \right>.
\end{equation}
and will analyze when such an assumption is representative of the colloid $\zeta$.

\section{Results}
\label{Results} 

We have performed simulations for a single spherical particle of radius $a$
embedded in cubic boxes with periodic boundary conditions, also varying the
concentration of added salt. In order to avoid finite-size effects,
for a given salt concentration we use a lattice length L at least three times
the corresponding Debye length, $\lambda_D$.  We have distributed uniformly the colloid charge
$Q$ on the lattice sites the colloid fills, and  choose $Q$   to correspond to
the predicted value of Oshima, according to eq.(\ref{Ohshima}). We will analyze
both a magnitude of $Q$ which  lies in the linear regime,  $\zetat_O = 1.0$,
and one for which the colloid is within the nonlinear regime,  $\zetat_O =
5.0$. We finally set the external electric field to $E=0.01$, well within
the linear regime $E \ll \frac{\zeta}{\lambda_D}$.

The values  for $\zeta$ obtained  from electrophoresis  are in principle different from those computed in equilibrium.  This is analogous to the
different measurements  for the size of a solvated particle, at
equilibrium or at hydrodynamic steady state (hydrodynamic
radius). We have therefore computed $\zetat_{MP}$ at thermodynamic equilibrium,
i.e. with equilibrated charges and  fluid at rest and dynamically, when a steady
state due to the external electric field is reached.  In all
simulations, our results show that the differences between the two values of $\zeta$ are smaller than the statistical precision. This similarity is expected  because of the small magnitude of the perturbing electric field and  colloid P\'eclet number, $Pe=\frac{a
v_{coll}}{D_{\pm}}$, where $v_{coll}$ is the velocity of the colloid at steady
state. For $Pe\geq 1$ the deformation of the salt cloud around the colloid induces significant deviations in the electrophoretic mobility[18]; we can hence envisage a corresponding departure of  the dynamic $\zeta$ from its equilibrium counterpart.
 Typically, as for simulations here, $Pe \ll 1$ and particle mobility
does not depend on $Pe$, hence the measured  $\zeta$ values at   equilibrium and
steady state are statistically indistinguishable.

As the only theoretical result available, $\zetat_O$, is
valid for a single particle in equilibrium at infinite dilution, we will 
compare  the simulation results  with eq.(\ref{Ohshima}) to assess the role of
the counterion concentration  at finite volume fraction. In addition, in order  to
analyze to which extent the midpoint stands as a proper definition for the
electrostatic radius, we run a pair of simulations for each colloid charge,
volume fraction and salt concentration used, changing the resolution of the
colloid in the lattice from low, using colloid radius $a=4.5$, to high $a =
9.0$, also doubling the corresponding box sizes (e.g. from $L=30$ at low
resolution to $L=60$ at high resolution for the highest volume fraction).

\begin{figure}
\centering
\mbox{\subfigure{\includegraphics[width=2.4in]{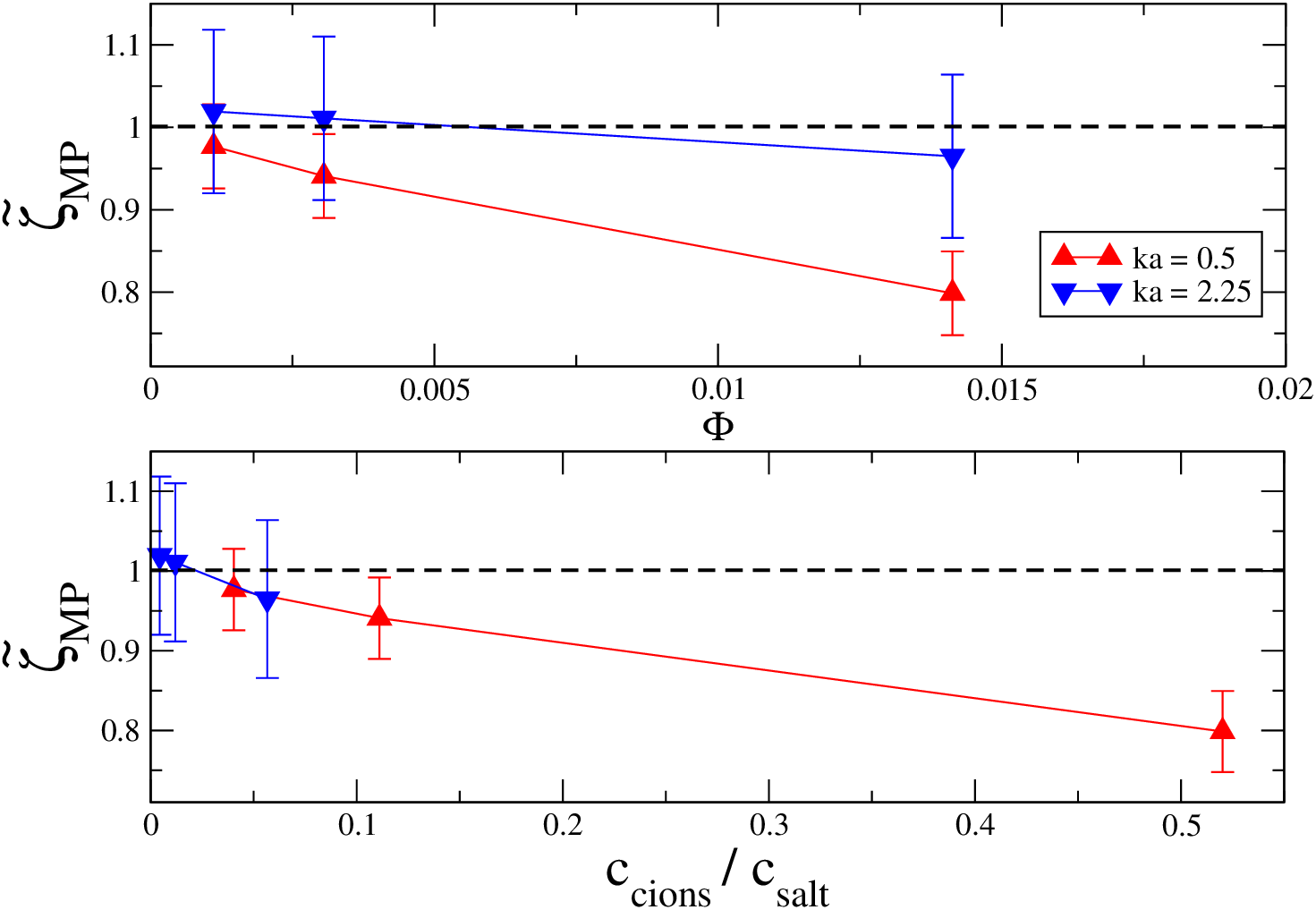}}\quad
  \subfigure{\includegraphics[width=2.4in]{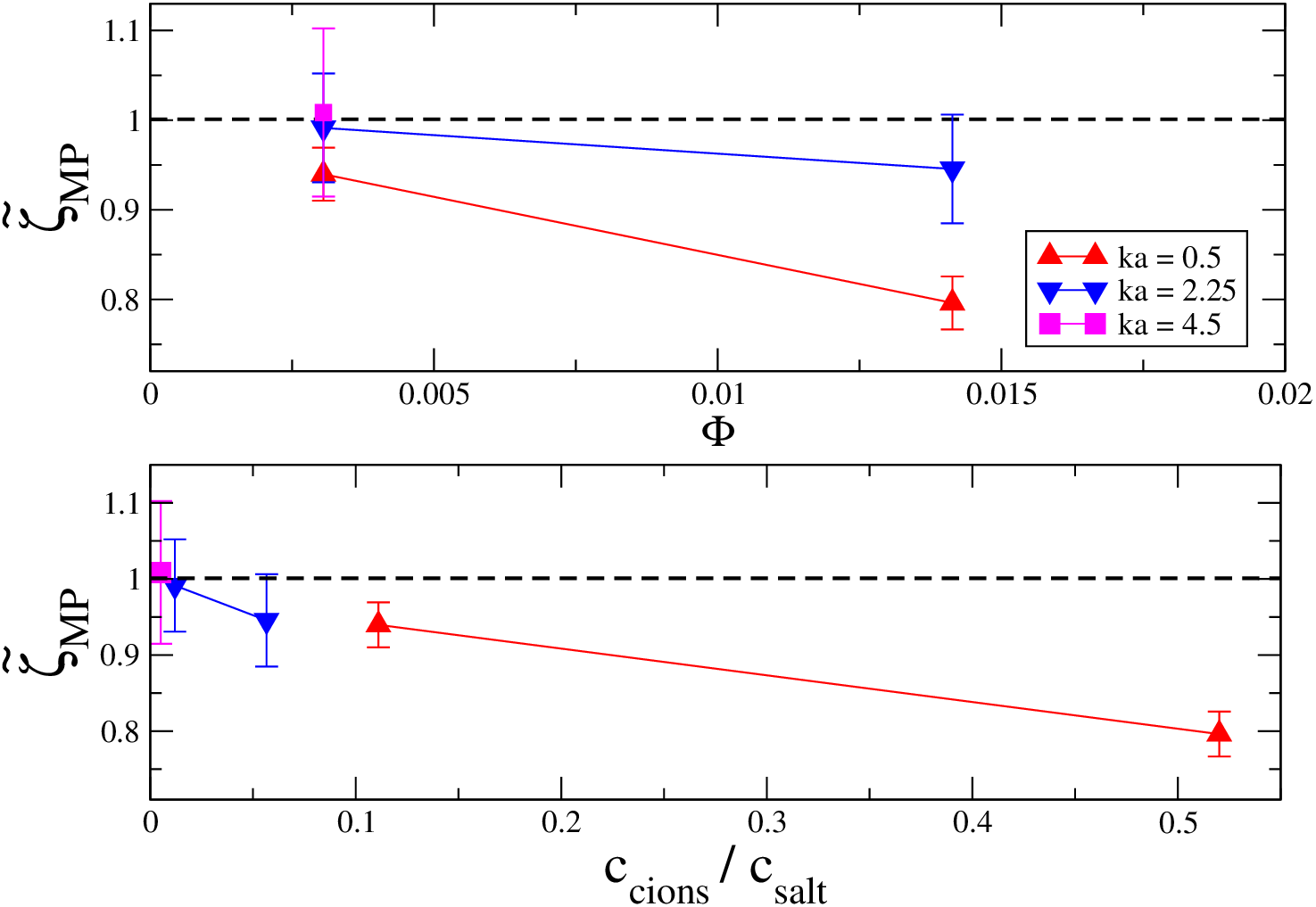} }}
  \caption{Left(Right): Electrostatic potential $\zetat_{MP}$ calculated halfway between
    boundary links (see text for definition) using a low(high)-resolution
      lattice. Colloidal charges for different salt concentrations ($ka=0.5,
          2.25, 4.5$) have been assigned so that the resulting theoretical zeta
      potential (eq.~(\ref{Ohshima})) is within the linear electrostatic,
                Debye-H\"uckel regime  ($\zetat_O = 1.0$, dashed line)}\label{fig1}
\end{figure}

Fig.\ref{fig1} displays $\zetat_{MP}$ for a weakly charged colloid
($\zetat_O=1.0$) as a function of the system size and salt concentrations,
where the left(right) panel corresponds to a low(high) colloid resolution.
The top panels show that the deviation of $\zetat$ from Oshima's prediction
varying $\Phi$ appears indirectly only through the corresponding change in
the  co- and counter-ion concentrations because the deviation decreases when
the salt concentration is increased. Therefore, according to
eq.~(\ref{DHeq2}), we can gain understanding by analyzing the dependence of
$\zetat_{MP}$ on the ratio between the  concentration of counterions released
by the particle, $c_{cions}$ (which decreases as $\Phi$ increases ) and the
concentration of salt $c_{salt}$ dissolved in the system (which is independent
of $\Phi$). The bottom panels  confirm an agreement within $10\%$ between
$\zetat_{MP}$ and $\zetat_O$ for $\frac{c_{cions}}{{c_{salt}}} \lesssim 0.1$.
In addition, comparing the  left and right panels, we conclude that in the linear
regime the lattice resolution does not play a significant role, as for a given
salt concentration  $\zetat_{MP}$ does not differ significantly and the  use of a more
computationally expensive refined lattice only reduces the error bars. In
addition, we note from bottom-rigth panel that discretization effects become more relevant when increasing the salt concentration. This sensitivity is associated to the reduction of  $\lambda_D$ and the corresponding loss of resolution in the electrostatic potential around the colloid.

\begin{figure}
\centering
\mbox{\subfigure{\includegraphics[width=2.4in]{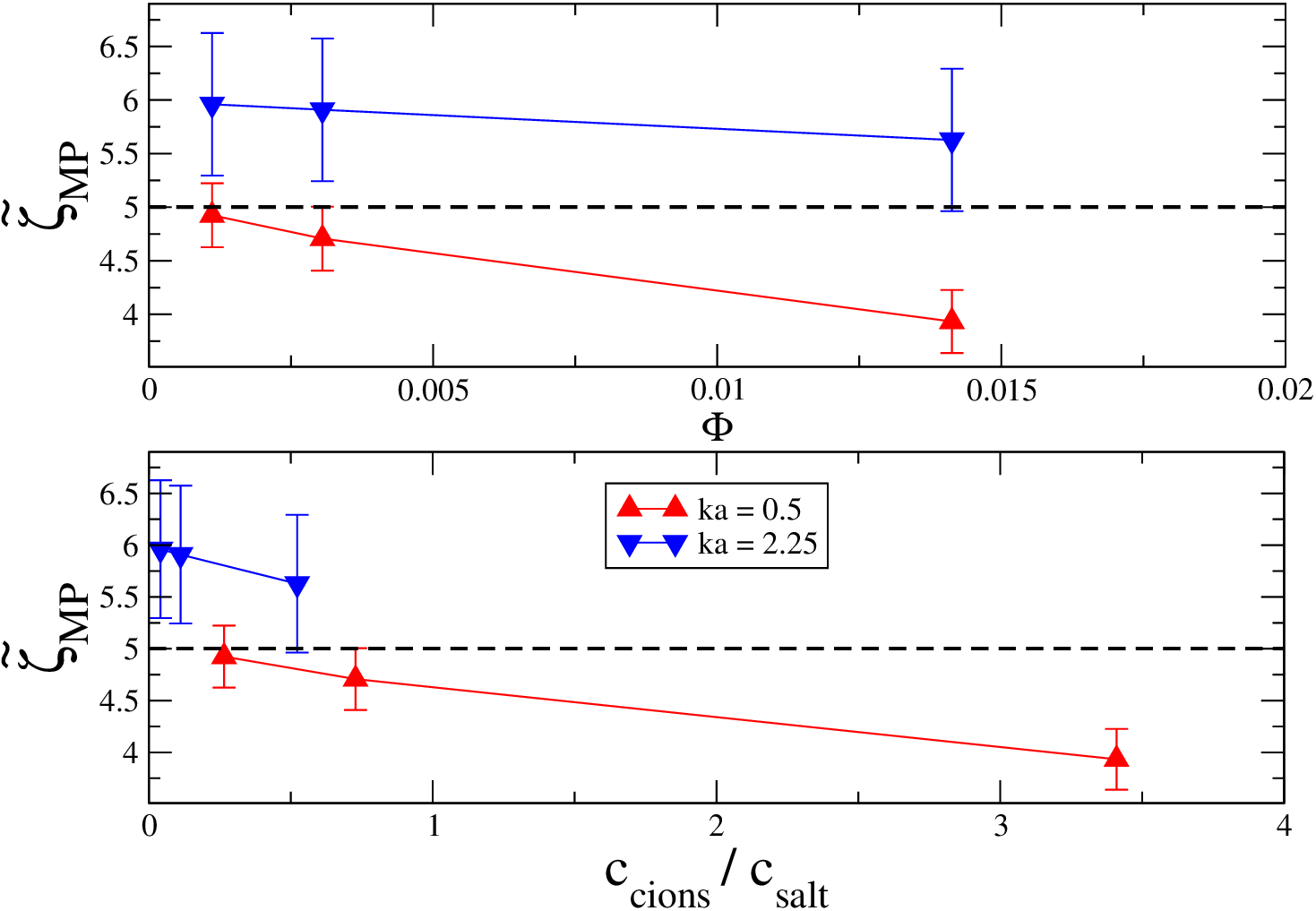}}\quad
  \subfigure{\includegraphics[width=2.4in]{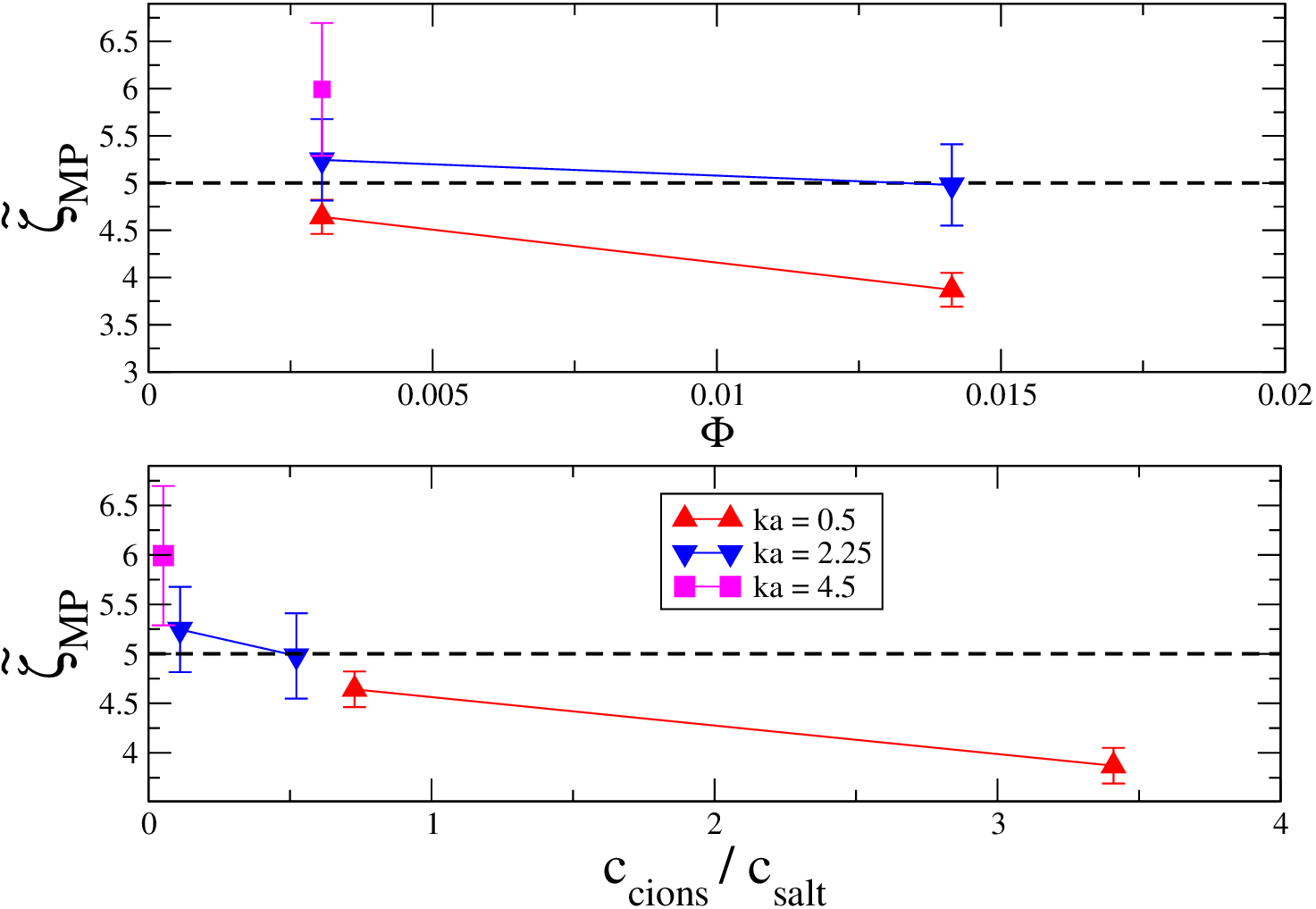} }}
  \caption{Left(Right): Electrostatic potential $\zetat_{MP}$ calculated halfway between
    boundary links (see text for definition) using a low(high)-resolution
      lattice. Colloidal charges for different salt concentrations ($ka=0.5,
          2.25, 4.5$) have been assigned to obtain a theoretical zeta
      potential (eq.~(\ref{Ohshima}), dashed line) $\zetat_O = 5.0$, within the nonlinear regime}
      \label{fig2}
\end{figure}

 Fig.~\ref{fig2}, organized analogously to Fig.~\ref{fig1}, shows results for a strongly charged colloid, $\zetat_O = 5.0$, for which the linearized  Debye-H\"uckel approximation does not hold.  We observe that the dependence of $\zetat_{MP}$ on $\Phi$ enters again indirectly through the relative changes in  $c_{cions}$, and that the convergence toward $\zetat_O$ can only be expected for high $c_{salt}$. However, as opposed to the linearized regime,  adding salt can lead up to a $20\%$ overestimation of $\zetat_{MP}$ for the highest salt concentration analyzed ($ka=4.5$), especially when using
a low resolution lattice (left panels).  In this strong coupling regime the electrostatic potential decays faster than $\lambda_D$[1]. This strong nonlinear behavior invalidates the midpoint as the natural choice for the electrostatic radius, which develops a dependence on the system parameters even when  the effects of finite $\Phi$ and $c_{cions}$ are negligible.

\begin{figure}
\centering
\mbox{\subfigure{\includegraphics[width=2.5in]{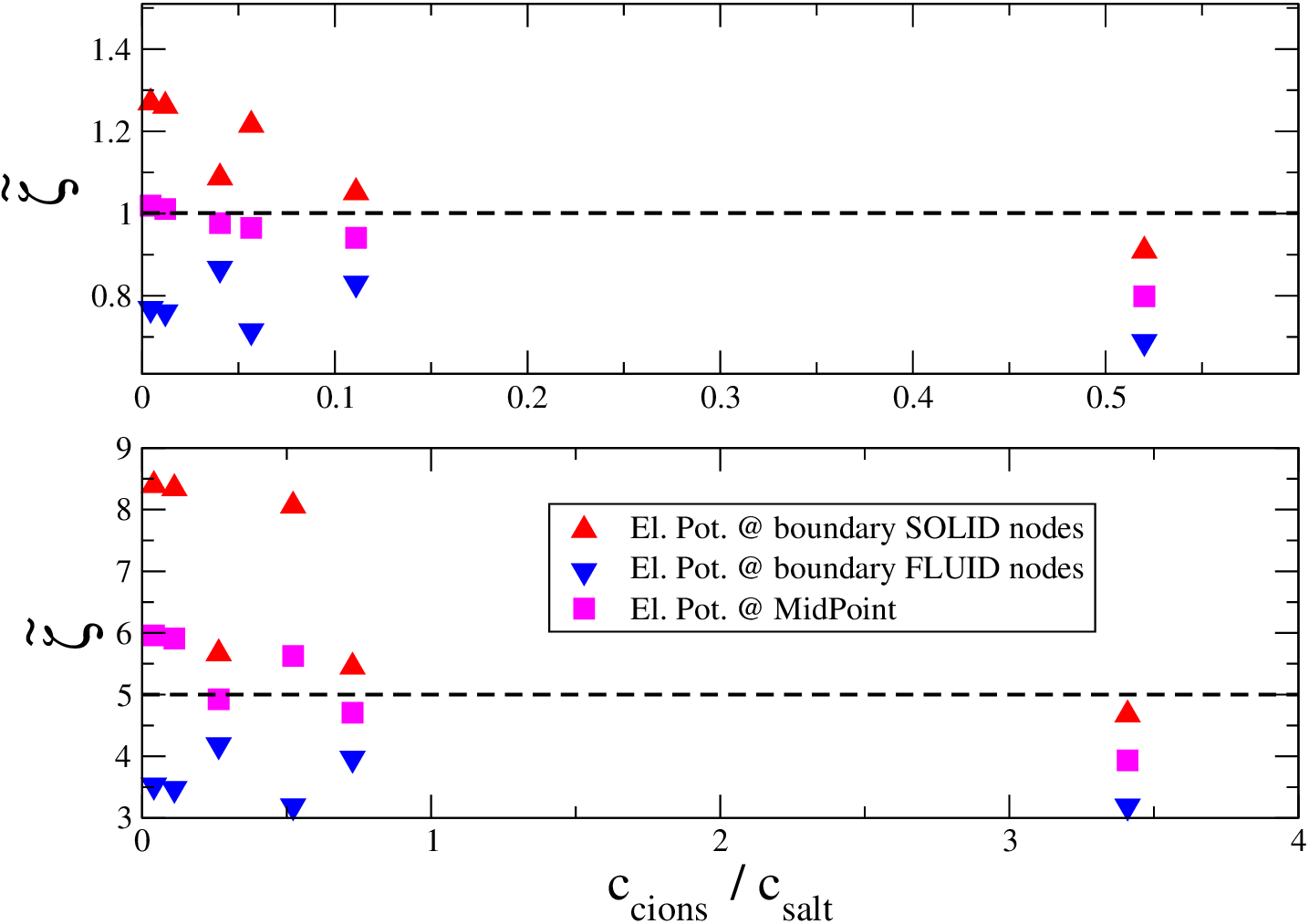}}\quad
  \subfigure{\includegraphics[width=2.5in]{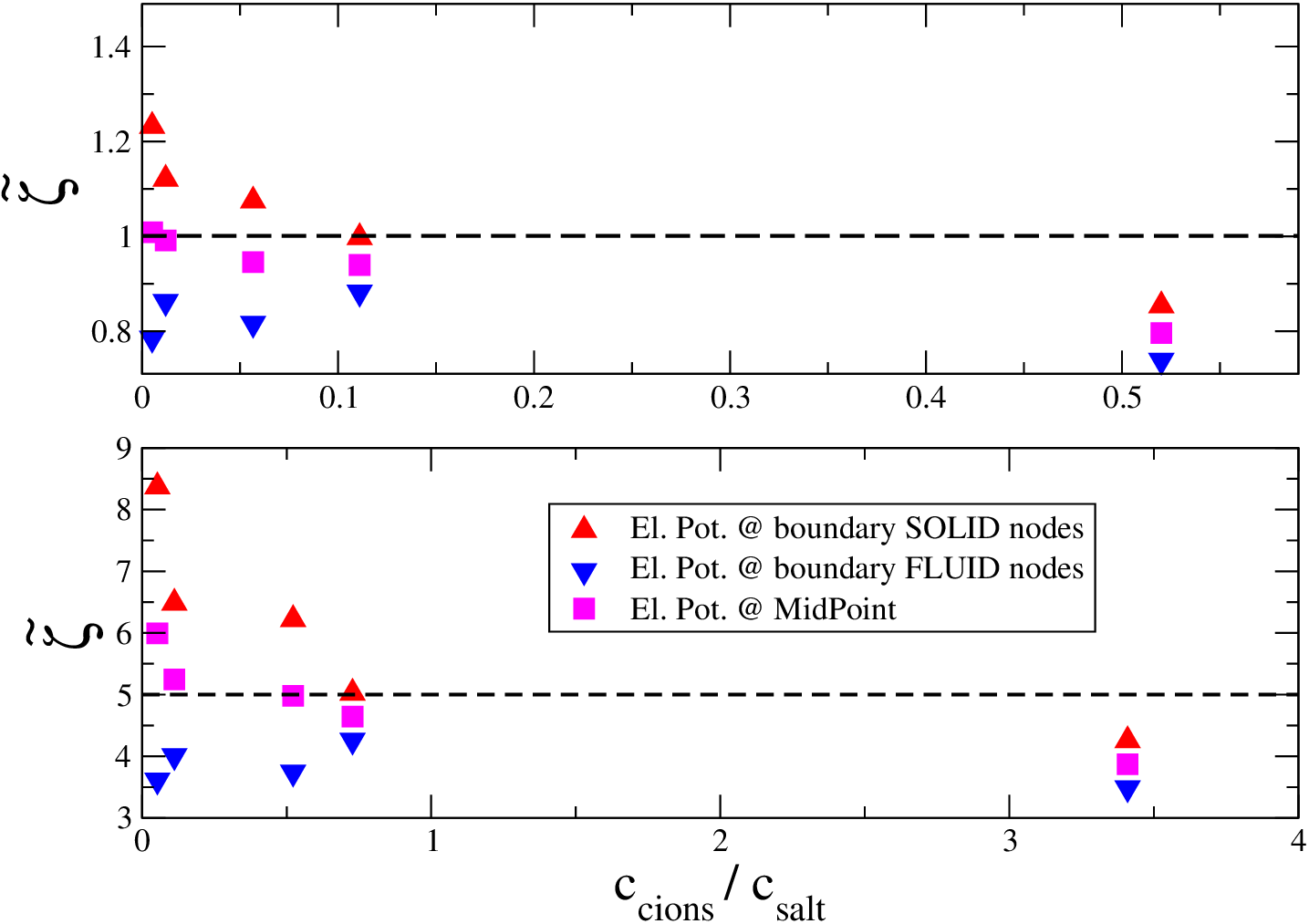} }}
  \caption{Left[Right]: Average electrostatic potential calculated using
    low[high] resolution lattice at midpoint (squares), solid (triangle-up) and
      fluid (triangle-down) nodes pertaining to boundary links in all
      simulation set-ups ($ka=0.5, 2.25, 4.5$, $\zetat_O = 1.0,5.0$). Top[Bottom] panels correspond
      to the case of a weakly [strongly] charged colloid.}\label{fig3}
\end{figure}

Fig.~\ref{fig3}  displays the measured electrostatic potential at the midpoint
between a solid and fluid node, $\zetat_{MP}$, together with the two extreme
situations where the electrostatic potential  is averaged over the solid and
the fluid nodes corresponding to the boundary links.  One can observe, as
expected, that $\zetat_{MP}$ always lies in between the other two estimates of
$\zetat$.  The differences observed between the left and right panels
indicate that the increase in resolution does not affect significantly
$\zetat_{MP}$, while  decreases the inaccuracy in the estimate of the
limiting values for $\zetat$. Generically, the better resolved the
colloidal particle (and the corresponding decay of the  electrostatic
    potential, $\varphi$) the less spreading between the different
electrostatic  potential estimates.  When we increase  $c_{cions}$, Oshima's
prediction does no longer hold and eq.~(\ref{Ohshima}) cannot be used to
identify the electrostatic radius as can be appreciated in the rightmost values
of $\zetat$ in the bottom panels.  However, the weak dependence of $\varphi$ in
the boundary fluid nodes suggests that  when $\zetat_{MP}$ is no longer valid,
we can still identify a  electrostatic radius slightly larger than the one
associated to $\zetat_{MP}$.  Analogously to the hydrodynamic radius, an
{\it ad hoc} calibration of the electrostatic radius at all salt concentrations
must be done to perform quantitative studies of the electrokinetics of
colloidal suspensions with LB.

\section{Conclusions}
\label{Conclusions} 
In this paper, we have analyzed  how to determine the electrostatic
potential at the surface of a particle, or zeta potential $\zeta$,  for
a colloidal suspension.  To this end, we have made use of a hybrid
mesoscopic  model which couples a discrete lattice formulation of
Boltzmann's kinetic equation for the solvent to a discrete solution of
the convection-diffusion equation for the charged ion species dissolved
in the fluid which implies treating counter- and salt ions as scalar
fields at the Poisson-Boltzmann (PB) level.  We have found that for
weakly perturbing electric fields, it is not possible to distinguish
between the equilibrium and dynamic  zeta potentials. We expect
differences will arise when the deformation of the charge layer around
the colloid becomes significant, a scenario which can be achieved, e.g.
when the P\'eclet number of the dissolved ions is not negligible. 

In particular, the comparison with Oshima's expression for $\zetat$ at
infinite dilution has allowed us to carry out quantitative  checks to
validate the code performance.  We have seen that the counterion
concentration has a significant effect on $\zetat$, leading to  a
decrease in its magnitude both in the linear and nonlinear regimes away
from Oshima's result. We have  shown that the ratio between counterion
and salt concentration controls the departure from Oshima's prediction,
and that its prediction works reasonably well for $c_{cions}/c_{salt} \lesssim
0.1$. We have also assessed the relevance of the lattice resolution and
have quantified its effects. We have seen that the mean of $\zetat$ over
the boundary nodes which determine the colloidal shape is in general a
good estimate of the electrostatic radius. However, for highly charged
colloids a more refined choice of the particle radius will be in general
needed. The effective radius will be slightly larger than predicted from
$\zetat_{MP}$ due to the nonlinear decay of the electrostatic potential
around the particle. This effective electrostatic radius, which needs to
be calibrated as a function of the salt concentration and particle
radius, will in general differ from the particle hydrodynamic radius and
requires a separate analysis. The overall dependence of the electrostatic
radius  both on the applied field and ion concentrations is weaker than
the one observed for the hydrodynamic radius. In most situations we
expect that an equilibrium calibration correcting from the finite
counterion concentration will provide a quantitative estimate for the
electrokinetics of colloidal suspensions.

\section{Acknowledgements}
G.G. acknowledges support from the IEF Marie Curie scheme. I.P.
acknowledges financial support from Direcci\'on General de
Investigaci\'on (Spain) and DURSI under projects  FIS\ 2008-04386 and
2009SGR-634, respectively.

\end{document}